\documentclass[10pt]{sigplanconf}

\usepackage[T1]{fontenc}
\newcommand{\tts}[1]{{\small \texttt{#1}}}

\def \Beta {\ensuremath{\mathsf{B}}}
\def \Pr {\ensuremath{\mathrm{Pr}}}
\def \E {\ensuremath{\mathrm{E}}}
\def \vF {\ensuremath{\mathbf{F}}}
\usepackage{balance}
\usepackage{times}
\usepackage{amsmath, amssymb}
\usepackage{url}
\usepackage{prettyref}
\usepackage{graphicx,xcolor}

\usepackage{hyperref}
\usepackage{url}
\usepackage{pgfplots}
\usepackage{prettyref}

\usepackage{amsthm}

\newrefformat{theoremm}{Theorem \ref{#1}}
\newrefformat{section}{Section \ref{#1}}
\newrefformat{eq}{Equation \eqref{#1}}
\def\xtheorem[#1][#2][#3]{\newtheorem{#2}[theoremm]{#3} \newrefformat{#2}{#3 \ref{#11}}}
\def\ytheorem[#1][#2][#3]{\newtheorem{#2}{#3} \newrefformat{#2}{#3 \ref{#11}}}
\xtheorem[#][theorem][Theorem]
\xtheorem[#][claim][Claim]
\xtheorem[#][lemma][Lemma]
\xtheorem[#][conjecture][Conjecture]
\xtheorem[#][proposition][Proposition]
\xtheorem[#][eg][Example]
\xtheorem[#][fact][Fact]
\xtheorem[#][corollary][Corollary]
\xtheorem[#][app][Appendix]
\xtheorem[#][alg][Algoritheorem]
\xtheorem[#][step][Step]
\theoremstyle{definition}
\xtheorem[#][definition][Definition]

\makeatletter
\def\url@leostyle{%
  \@ifundefined{selectfont}{\def\UrlFont{\sf}}{\def\UrlFont{\small\bf\ttfamily}}}
\makeatother
\newrefformat{sec}{Section~\ref{#1}}
\newrefformat{fig}{Figure~\ref{#1}}

\begin{document}

\setlength{\pdfpageheight}{\paperheight}
\setlength{\pdfpagewidth}{\paperwidth}

\conferenceinfo{PLATEAU '15}{Month d--d, 20yy, City, ST, Country}
\copyrightyear{2015}
\copyrightdata{978-1-nnnn-nnnn-n/yy/mm}
\doi{nnnnnnn.nnnnnnn}

\title{Frequency Distribution of Error Messages}

\authorinfo{David Pritchard}
           {Center for Education in Math and Computing, University of Waterloo, Canada\thanks{Work started while located at Princeton University and completed at U.~Southern California. Currently located at Google Los Angeles.}}
           {dagpritchard@uwaterloo.ca}

\maketitle
\begin{abstract}
Which programming error messages are the most common? We investigate this question, motivated by writing error explanations for novices. We consider large data sets in Python and Java that include both syntax and run-time errors. In both data sets, after grouping essentially identical messages, the error message frequencies empirically resemble Zipf-Mandelbrot distributions. We use a maximum-likelihood approach to fit the distribution parameters. This gives one possible way to contrast languages or compilers quantitatively.
\end{abstract}

\category{D.3.4.}{Programming Language Processors}{Compilers, Run-time environments}

\keywords
	Error messages, empirical analysis, usability, education.

\section{Introduction}
This work started as an offshoot of Computer Science Circles (CS Circles) \cite{PV13,PGV15}, a website with 30 lessons and 100 exercises teaching introductory programming in Python. It contains a system where students can ask for help if they are stuck on a programming exercise. Often, students reported being stuck because they could not comprehend an error message, asking for a better explanation of what the compiler/runtime was trying to say. E.g., the message
$$\texttt{\small SyntaxError: can't assign to function call}$$
might not be understood by a novice who wrote \verb|sqrt(y)=x|.

Motivated by this, we decided to systematically improve the error messages that students received. There is copious literature on writing good error messages \cite{MM67,H74,S82,T10,MFK11a,MFK11b}, but how can this advice be incorporated into the programming ecosystem? One approach would be making upstream improvements to the compiler/runtime, but this can take a long time, and not all audiences would appreciate the changes that would most benefit novices. A second approach would be to write a tool that analyzes code from scratch, looking for common syntactic bugs or likely semantic mistakes. The literature includes many such tools: see \verb|checkstyle|, \verb|findbugs| and \cite{J85,S95,L02,HMRM03,FCJ04,LFGC07}.

We chose a more lightweight approach: augmenting the normal error messages with additional explanations. To wit, we compile and execute the code as usual, and then add a beginner-appropriate elaboration of the resulting error message, implemented by rendering the normal error with a clickable pop-up link to the explanation.
This augmenting-explanation approach has been previously used on a small scale with Java compiler errors \cite[\S 5.2]{Coull08}, Python runtime errors \cite[\S 5.2.1]{Hartz12}, and C++ STL compiler errors \cite{Zolman05}.

It has long been observed that ``a few types of errors account for most occurrences'' \cite{RD78}, see also \cite{Denny12}. In order to make sure that a small number of explanations would be useful as often as possible, we had to answer the following question: \emph{what error messages are the most common}? Counting error message frequencies has a long history, starting from assembler \cite{MM67,CH76} and \verb|SP/k| \cite{H74}, with renewed interest more recently, using much larger data sets \cite{JCC05,Jadud05,Denny12,BKMU14,SSEAB14}.

Using with the history of all previous submissions, we determined the essentially distinct error messages and their frequencies (see Section~\ref{sec:data}), available online at \url{http://daveagp.github.io/errors}. We wrote explanations for the 36 most common messages. Regular expressions were used to aid the implementation. At the most basic level, some errors were made more readable by elaborating them into a full paragraph of text rather than a one-line message. Some explanations include concrete examples of code that causes the same error message, and a description of how to fix it. See \cite{MM67,H74,S82,T10,MFK11a,MFK11b} for advice on writing error messages. 
Given a large data set, the work involved in this group-and-explain approach is modest and not technically challenging, so we would recommend it in any beginner-facing system. Moreover, in internationalized settings, one can then add explanations in other languages (this has been implemented in CS Circles' Lithuanian translation).


This paper compares and contrasts the most common error messages in CS Circles with those in another programming language. The Blackbox project~\cite{KU12,BKMU14} is a large-scale data collection-and-sharing project using BlueJ, a Java programming environment oriented at beginners. We obtained the error messages from all recorded compilation and execution events, grouping and counting the essentially different messages like we did for the Python data set. Comparing the two data sets, we found that both error message frequency distributions resembled the same family of distributions, the Zipf-Mandelbrot distribution~\cite{Mandelbrot53}. For these data sets, this means that for any integer $k,$ the frequency of the $k$th most common error is approximately proportional to $1/(k+t)^\gamma$ where $t$ and $\gamma$ are parameters of the data set. In order to determine the best values for these parameters, we propose using a simple maximum-likelihood approach.

\subsection{Discussion and Other Related Work}\label{sec:related}

Orthogonal to purely quantitative analysis, a large body of work focuses on manual categorization of errors. This allows researchers to get more accurate results, and to precisely understand the psychological state of the user, rather than focus on the compiler-generated error messages themselves. Good reasons for doing this include that ``A single error may, in different context, produce different diagnostic messages'' and that ``The same diagnostic message may be produced by entirely different and distinct errors,'' see \cite{McCall14}. This analysis also helps measure whether a compiler's error messages are appropriate (e.g., see~\cite{RD78,Johnson90}). This analysis is  important for compiler designers, language designers and educational research, but it is not our focus. 

The comparison of error message frequencies between different languages raises many interesting open-ended questions. Even within the same language, some compilers are significantly better or worse than others; see Brown's amusing crowdsourcing of Pascal error messages~\cite{Brown82,Brown83} as well as \cite{NPM08,T10}. One way to view different error message distributions is to imagine the extremes: the worst possible language would only ever say ``?'' without elaborating (this has been formally evaluated, see~\cite{S82}), while the best possible language would, like a human tutor, always give a perfectly adapted explanation. The exponent $\gamma$ in our work is one way to measure where a language sits between these extremes. However, simpler measures such as entropy could also be used. Also, a single quantitative measure should not be treated as paramount without context. When comparing languages/compilers (e.g., \cite{McIver00}), statistical fitness is less important than overall usability, including measures like time between errors and time to achieve user goals.

A notable alternative approach to improving student feedback based on large-scale data, rather than focusing on error messages, is the HelpMeOut system \cite{HMBK10}, which uses a detailed repository of past student work sessions to find old errors similar to new ones and make suggestions of how to fix them.

To our knowledge, this paper is the first one to examine any link between programming error messages and statistical distributions. The special case $t=0$ of the Zipf-Mandelbrot distribution is known as the power law distribution. It arises empirically in data sets such as the frequency of distinct words in books, of links in webpages, and of citations in literature. Caveats apply here~\cite{CSN09,GMY04,Newman05}, including: that generative explanations of how these distributions could arise are tenuous; that near-power-law data sets may be even closer still to other distributions; and that analyzing such data sets has common pitfalls like using linear regression. Another caveat for our work is that the distributions of error messages will depend on the nature of the users, and the kind of setting in which the work is collected. In our case, both data sets come from a very large, open project intended for beginners. We anticipate that a data set where students only work on a fixed set of exercises could be skewed in some way, but both BlueJ and the CS Circles ``console'' allow students to do any sort of open-ended programming.

See \cite{PNFB05} for discussion of power laws in runtime object-reference graphs of industry-scale computer programs.



\section{Data Sets}\label{sec:data}
Our first data set is the Python corpus from CS Circles. Amongst the first 1.6 million code submissions, 
about 640000 resulted in an error.
Our second data set is the Java corpus from BlueJ Blackbox. We specifically considered 
the ``compile'' events, of which there were about 8 million, half of which produced an error, and the ``invoke'' events, of which there were about 5 million, about 260000 of which produced a syntax error and 180000 of which produced a run-time error. We did not include the codepad or unit test events, both of which are an order of magnitude smaller. 



In both cases, following \cite{NPM08}, we only counted the first error message. This tends to be the most accurate error (since a syntax error can cause new valid parts of the program below to be reported as errors) and it is also the error that the programmer is most likely to pay attention to and fix first. Moreover, CS Circles only shows the first error message in its user interface; and even for a UI like BlueJ that shows multiple errors, beginner students often (by habit or by instruction) fix only one at a time and then recompile/re-run.\footnote{It would not be invalid to investigate data sets where all errors are reported and counted, but a worry is that it might say more about the statistics of chain effects in syntax errors and less about the actual underlying bugs. Two other strategies, ``count-all'' and ``count-distinct,'' are used in~\cite{SSEAB14}, though their study participants were professionals and not novices.}


After obtaining these raw data sets of hundreds of thousands of error messages, we had to count how many time each distinct message occurred.
It is necessary to ``sanitize'' the data by removing parts that pertained to specifics of user code rather than the kind of error. For instance,
\tts{NameError: name 'x' is not defined} should be understood by our system to be essentially the same error as \tts{NameError: name 'sum' is not defined} so that the same explanation will appear in either case. The sanitization was an iterative process. Simple heuristics handled most cases correctly, and in total we needed about 20 sanitization rules for Python and 50 for Java, implemented using regular expressions.

There is a question of how far one should sanitize. Should these two error messages be considered the same?
{\small \begin{verbatim}
RuntimeError: maximum recursion depth exceeded
   while getting the repr of a list
RuntimeError: maximum recursion depth exceeded
   while getting the repr of a tuple
\end{verbatim}}
Overall we tended to use fewer sanitization rules rather than more (considering the above to be different); a similar approach was used in \cite{SSEAB14}. Conceptually, to fix a single objective goal for sanitization, we imagined that each category should uniquely correspond to a single line of source code of the compiler/runtime where the error is first detected.

Another step in sanitization was to remove any non-English error messages, to avoid inadvertently seeing the same patterns repeated in multiple languages, which might affect the results. This was done by removing all messages with non-ASCII characters, and manual filtering.




\subsection{Overview of Data Sets}\label{sec:datasets}
The Python data set yielded 309710 syntax errors and 333538 compile-time errors. The Java data set yielded 4002822 compile-time errors and 129650 run-time errors.  
Note that the Java data set has a much smaller proportion of run-time errors than Python (only about 3\% rather than almost half). But to a degree, this difference is inherent in the language, since many errors that would occur at compile-type in Java's strict typing-and-scoping system are not encountered until run-time in Python. 

After sanitization and grouping, the Python data set yielded 283 distinct error messages. Of these, 17 occurred exactly twice and 42 occurred only once (for example, ~\tts{ValueError: Format specifier missing precision} and \tts{SyntaxError: can't assign to Ellipsis}). The Java data set yielded 572 error messages in total; 65 occurred exactly twice and 127 occurred only once (for example, \tts{com.vmware.vim25.InvalidArgument} and \tts{cannot \\ create array with type arguments}).

Errors are not completely parallel for both languages. For example, Java allows function overloading, i.e.~two functions with distinct signatures but the same name. In Python, this must instead be implemented by a single function that takes different actions depending on the runtime number and type of its argument(s). It is the function's responsibility to generate the error message. It turns out that not all such functions generate identical messages and so the single Java error message \verb|no suitable method found| corresponds to more than one distinct Python error message:
\\ {\small \texttt{f~argument~must~be~a~string~or~number, not T}}
\\ and {\small \texttt{f~arg~1~must~be~a~type~or~tuple~of~types}}.

\eject

\begin{figure}[tb]
\centering
\includegraphics[width=8cm]{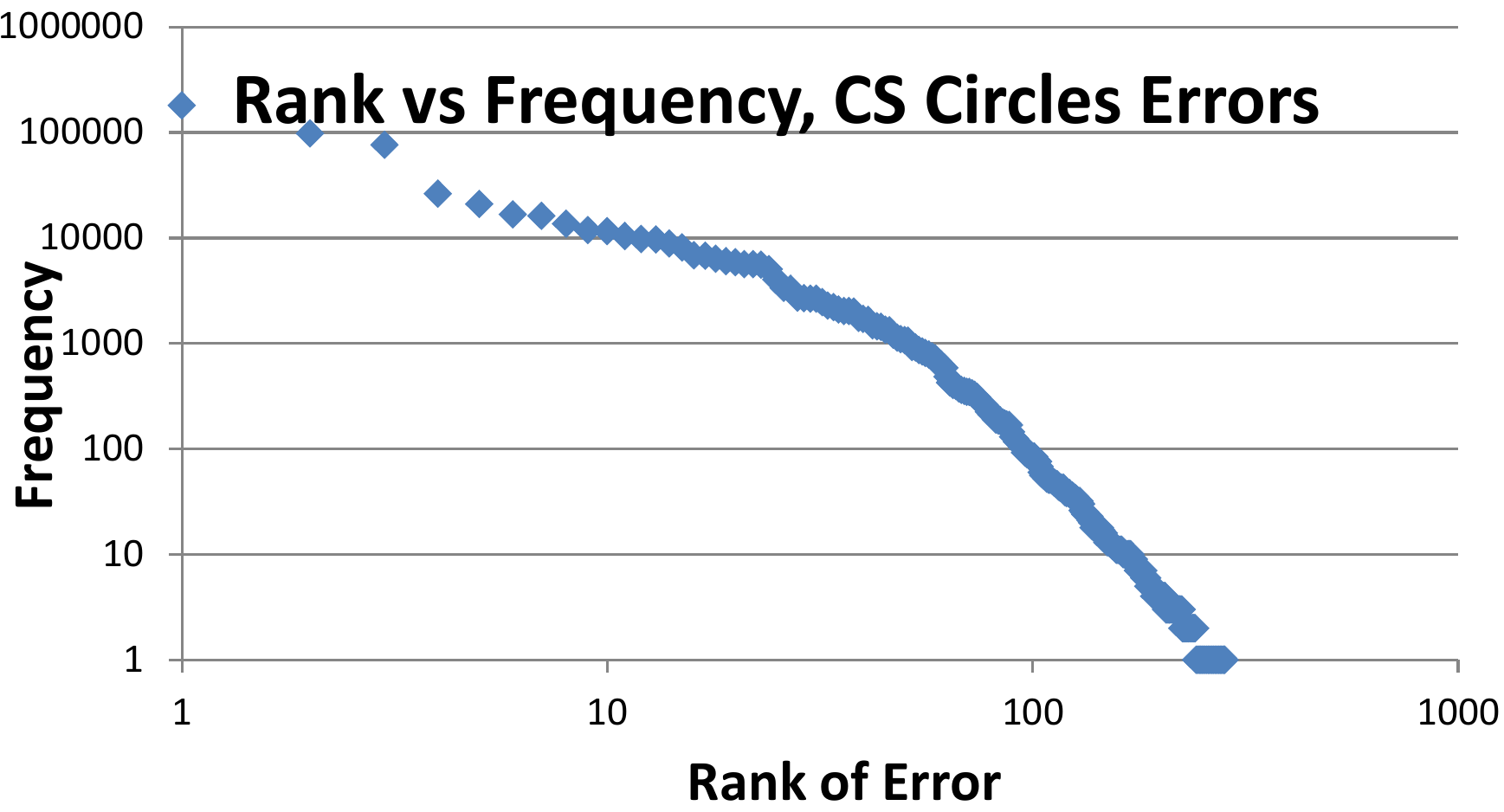}
\includegraphics[width=8cm]{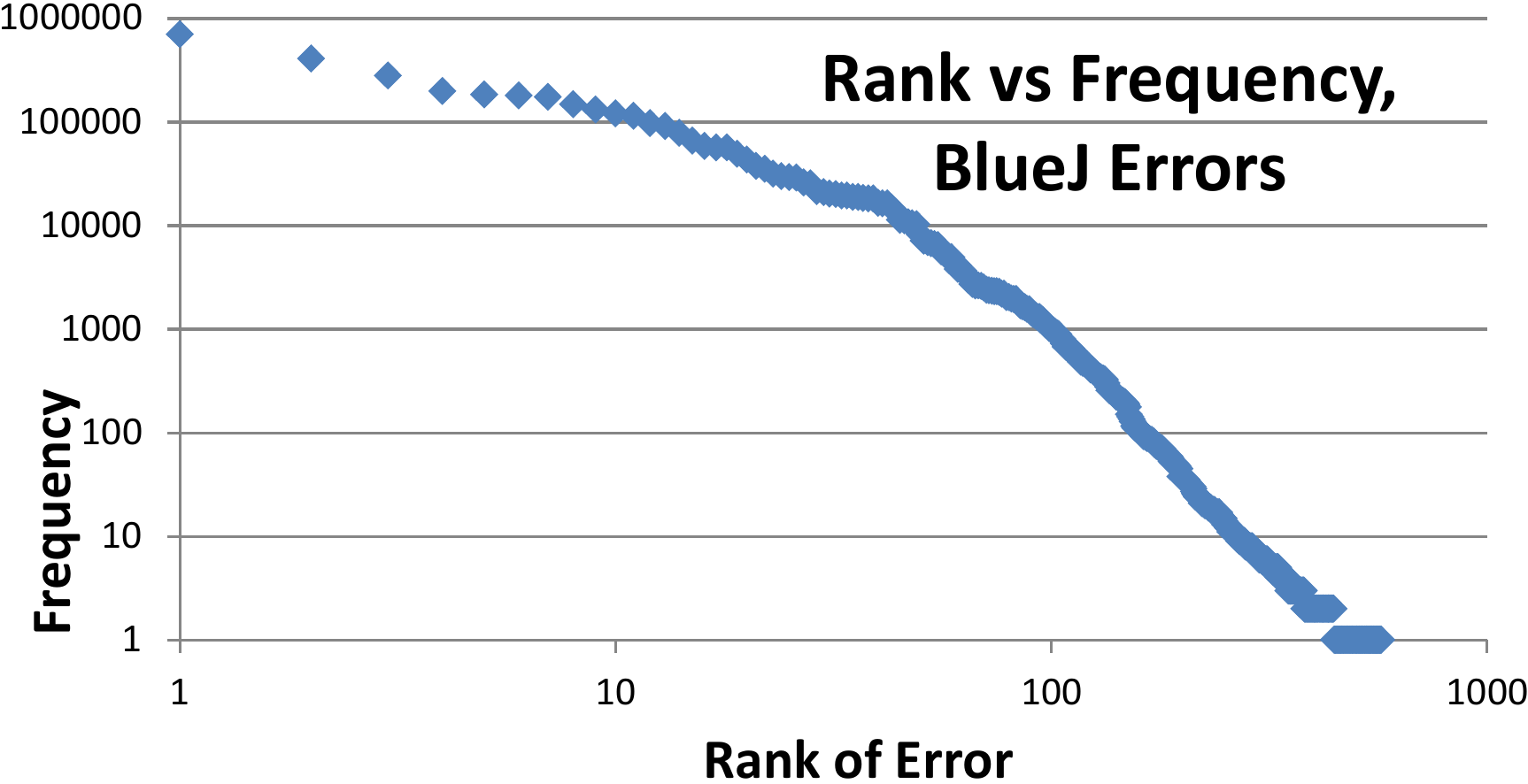}
\caption{The two data sets for our study. The CS Circles data set is Python, while BlueJ is Java. The plots are log-log.}
\label{fig:rawdata}
\end{figure}

The 5 most common Python errors were:
{\small
\begin{verbatim}
179624 SyntaxError: invalid syntax
 97186 NameError: name 'NAME' is not defined
 76026 EOFError: EOF when reading a line
 26097 SyntaxError: unexpected EOF while parsing
 20758 IndentationError: unindent does not match
       any outer indentation level
\end{verbatim}}

The 5 most common Java errors were:
{\small
\begin{verbatim}
702102 cannot find symbol - variable NAME
407776 ';' expected
280874 cannot find symbol - method NAME
197213 cannot find symbol - class NAME
183908 incompatible types
\end{verbatim}}


We plot both data sets in Figure~\ref{fig:rawdata}. The $x$-axis measures the rank of each error message (with 1 being the most frequent) and the $y$-axis measures the number of times each error occurred. Using a logarithmic scale is necessary for the changes in the $y$-axis to be visible, and we also use a logarithmic scale for the $x$-axis. Notice that both data sets give rise to similar distributions; in the rest of the paper we will try to describe them in a common framework.

\subsection{Notation}
For any given data set, we will use $N$ to denote the total number of errors logged, and $M$ for the number of distinct error types. For example, the Python data set has  $N=643248$ and $M=283$. Let $F_k$ denote the number of times that the $k$th-most common error occurred, e.g.~$F_1 = 179624$ for Python. We will also write $F_{\max} := F_1$ as an alternate symbol for the same value, when we wish to emphasize that it is the maximum frequency. The smallest frequency $F_M$ is 1 for both of our data sets. In lexicography, the items occurring just once are known as the \emph{hapax legomena} of the corpus. An \emph{$f$-legomenon} is any error message that appears exactly $f$ times. We will use the symbol
$$\#F^{-1}(f)$$
to denote the number of $f$-legomena. The first few counts of $f$-legomena in our data sets is listed in Table~\ref{table:legomena}.
\begin{table}
$$
\begin{array}{|c|c|c|c|c|c|c|}
\hline
\#F^{-1}(f) \textrm{ where } f \textrm{ is:}& 1 & 2 & 3 & 4 & 5 & 6 \\
\hline
\mathrm{Python}& 42 & 17 & 19 & 13 & 6 & 4\\
\hline
\mathrm{Java}& 127 & 65 & 31 & 17 & 18 & 15\\
\hline
\end{array}
$$
\caption{Number of $f$-legomena in each data set.}
\label{table:legomena}
\end{table}




\section{Power Law Distributions}
When studying frequency counts of different objects, a discrete \emph{power law} distribution is one in which the frequency $F_k$ of the $k$th most common item is proportional to $1/k^\gamma$. As mentioned in the introduction, power law distributions provide good fits to many unrelated empirical distributions. A common example is that in the novel \emph{Moby Dick}, the frequency of the $k$th-commonest word is approximately proportional to $1/k^{1.05}$. ``Zipf's law'' is sometimes used as a synonym for the discrete power law, but sometimes also refers to the special case $F_k \propto 1/k$ where $\gamma=1$. There is also a large body of work on continuous power laws, where one sorts items by some magnitude that takes on continuous values, and examines the relationship between rank and magnitude. See many examples of both types in \cite{CSN09}.

Note that sanitization of error messages is particularly important because of the fact that many natural languages follow power-law curves. If we did absolutely no sanitization, then our error message distributions would have significant aspects determined by the frequency distribution of variable names chosen by users, and finding a power law describing the latter would be less surprising, given that natural language is already known to exhibit power law behaviour.

We now turn to analyzing our data sets from the power law perspective. Do they approximately satisfy a power law? This did not appear to be the case: a power law, when plotted on a log-log scale, should give a straight line, but it is clear from Figure~\ref{fig:rawdata} that this is not an accurate description of our data set.

\subsection{Zipf-Mandelbrot Distributions}\label{sec:zmd}
\begin{figure}[b]
\centering
\includegraphics[width=8cm]{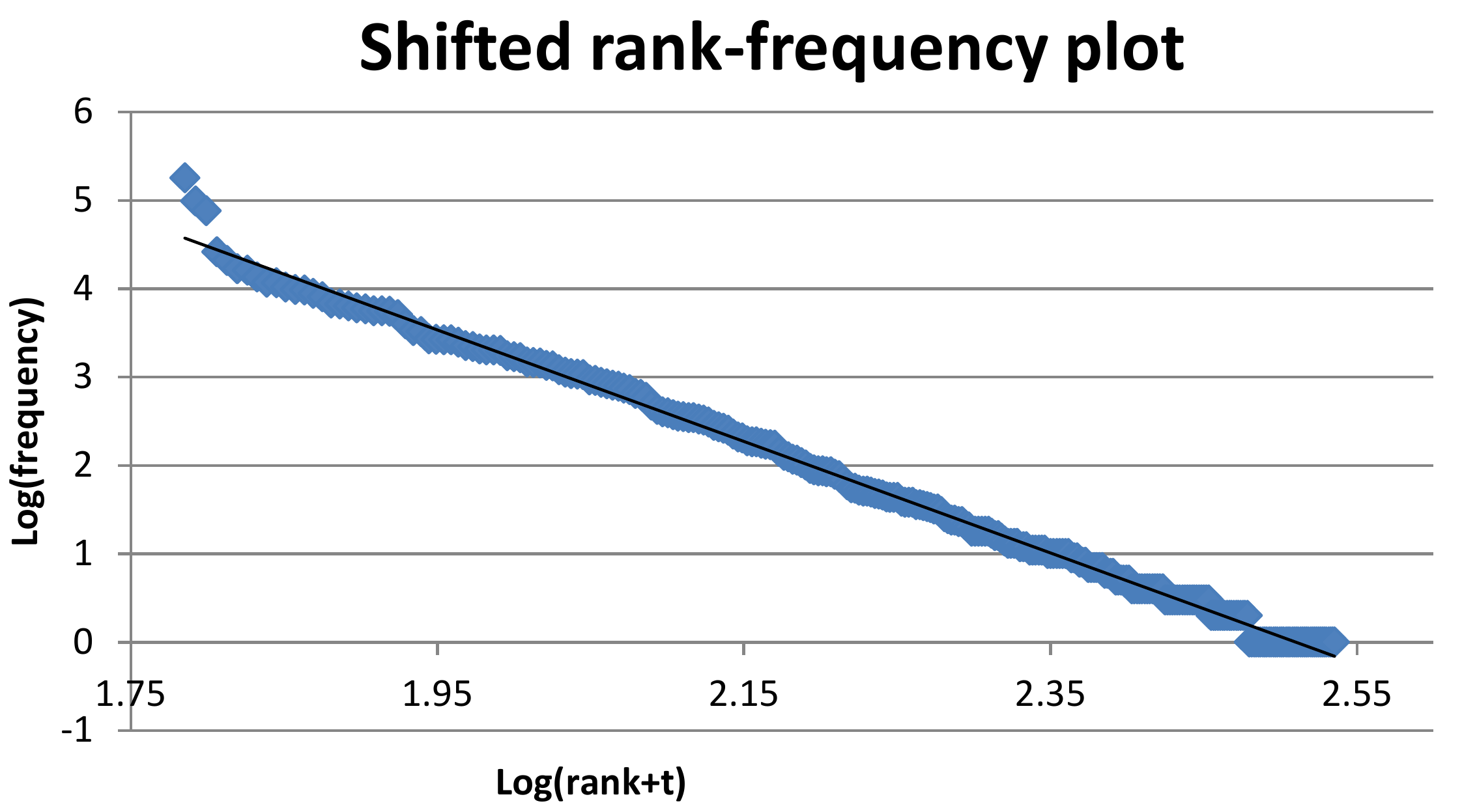}
\caption{
The Python data set with the (pre-logarithmic) $x$-axis shifted by $t=60$, and a straight line with slope $-\gamma = -6.3$ that approximately fits most of the data.}
\label{fig:rf}
\end{figure}

There is a generalization of power law distributions called the \emph{Zipf-Mandelbrot} family of distributions. These distributions are defined, using two parameters $t$ and $\gamma$, by
$$F_k \propto \frac{1}{(k+t)^\gamma}.$$
Such a sequence should appear linear on a log-log plot provided that along the $x$-axis, we plot the logarithmic positions of $(k+t)$ rather than of $k$. When we tested plotting these distributions in this modified way, for an appropriate value of $t$, we obtained a much more persuasive fit: in Figure~\ref{fig:rf}, which has the shift $t=60$, the points very nearly fall on a line. This shift was obtained by trial-and-error, and the line drawn in has slope $-\gamma = -6.3$. In the rest of the paper we aim to give a more principled way of estimating $t$ and $\gamma$.

Is a Zipf-Mandelbrot distribution plausible? Here is one argument that, if we accept that power laws can arise in natural settings, that there is reason to suspect that Zipf-Mandelbrot laws can too. It is not meant to give an exhaustive explanation, just an argument for plausibility. Suppose we start with a power law, and then coalesce several items together. I.e., replace several distinct error messages with a single unified message having the sum of their frequencies. (In a list of English words, the analogy would be that a single word has multiple meanings.) The effects of this message-merging would be twofold: the resulting new message would be an outlier to the original power law curve; and the remaining data points, when plotted on a rank-frequency scale, would be shifted several positions to the left, i.e.~they would follow a Zipf-Mandelbrot distribution instead of a power law distribution. This is indeed a plausible scenario for the Python data set! The most common error, \verb|SyntaxError: invalid syntax|, is very generic. It can be obtained by writing two tokens in a row (such as forgetting a comma or quote marks), using an assignment statement in place of a conditional expression (such as using \verb|if a=b:| instead of using \verb|==|), by mismatching parentheses, etc. 

\subsection{Consequences of a Zipf-Mandelbrot model}
What behaviour does a Zipf-Mandelbrot model predict? It postulates that there is some innate ordering of error messages, from most frequent to least frequent, so that the inherent probability $F^*_\ell$ of the $\ell$th most frequent error message is proportional to $(\ell+t)^{-\gamma}$. The reason that we use the subscript $\ell$ here is that our concrete data set arrives via sampling from the inherent distribution. So like a sampling error, the observed ordering of messages from most to least frequent is not necessarily the exact same as the innate ordering.


An interesting aspect of this model is that it assumes $F_\ell^* \propto (\ell+t)^{-\gamma}$ continues to hold for arbitrarily large $\ell$. Can this be plausible: is the total number of possible errors infinite? We will accept this as a reasonable hypothesis, which if not literally true, could continue long enough to be consistent with the size of any measured data set, for the following reason, using Python as an example.  The most common errors we see are the ones in the Python core code (the syntax errors, and runtime errors from the ``builtins''). Less frequently we start to see errors from Python modules, such \tts{ValueError: math domain error} within the \verb|sqrt| function of the \verb|math| module. While this is the only module taught on the site, users have occasionally submitted code using other common modules like \verb|time|, \verb|random| and \verb|functools|, each of which comes with its own specific errors. Moving on, there would be errors from rarely-used modules, then even after this, modules that users may with more or less frequency import (or copy in) themselves. For example, we observed a \tts{mainfile: error: must provide name of pdb file} error caused by someone who copied in a Python program for use with the X-PLOR biomolecular structure determination software~\cite{SKTM03}. The same phenomenon happens in the Java data set. So errors with arbitrarily small inherent frequency are not unreasonable despite the finite size of the languages.

\section{Analysis}
To fit our data to a Zipf-Mandelbrot distribution, several approaches are possible. For power laws, the naive approach, using a least-squares fit to a linear log-log plot (c.f.~Figure~\ref{fig:rf}) is known to introduce errors~\cite{GMY04}. Rather, we will follow Newman~\cite{Newman05}, who considered maximum likelihood estimation methods for power laws. Some work will be needed to extend this to Zipf-Mandelbrot distributions.

The method in \cite{Newman05} involves a particular way of processing the data; let us mention the motivation. The direct approach to maximum likelihood estimation would be to determine the $\gamma$ and $t$ that maximize $\prod_k (C/(k+t)^{\gamma})^{F_k}$ where $C$ is the normalizing constant with $C \cdot \sum_{k=1}^\infty (k+t)^{-\gamma} = 1$. Izs{\'a}k~\cite{Izsak06} suggests this. But trying this approach gives unsatisfactory results with any of our data sets --- the curves produced fit the data very poorly except in the regime of $F_1$. The calculation goes wrong because it is too heavily biased by the highest-frequency errors. (For Zipf-Mandelbrot in particular, if the argument in Section~\ref{sec:zmd} were to be true, then it should be no surprise that fitting to $F_1$ would be problematic, since $F_1$ would be an outlier from the norm.)
Also, the most likely fit entails that the innate order exactly matches the observed frequency-ordering of error messages~\cite{Izsak06}, which is itself unlikely.

\subsection{Probabilities of Frequencies}
This motivates the maximum likelihood method on frequencies~\cite{Newman05,CSN09}. It starts by taking a different view of the data set. Using the Python data set as a concrete example, we imagine the frequency vector $\vF = (179624, 97186, \dotsc, 1, 1)$ itself as being an unordered set of $M$ data points from a parameterized distribution --- given a new error message, how frequent is it? This distribution-on-frequencies is a transformed version of the inherent distribution $F^*$, and also depends on the data set size. The goal, then, is to choose the parameters so as to maximize the likelihood of observing $\vF$.

The analysis in~\cite{Newman05,CSN09} primarily achieves rigor for continuous distributions. For discrete distributions, it turns out that the distribution-on-frequencies is actually given by another distribution which seems to have been first studied by Evert~\cite{Evert04}. To describe it we recall the $\Gamma$ function, which is the (shifted) analytic continuation of the factorial function, satisfying $\Gamma(n)=(n-1)!$ at positive integer values and $\Gamma(n)=(n-1)\Gamma(n-1)$ on its whole domain. The beta function, another standard function, is a continuous analogue of the binomial coefficient, defined by
$$\Beta(x, y) := \Gamma(x)\Gamma(y)/\Gamma(x+y).$$

Then, finally, the \emph{Evert distribution} is the frequency distribution, parameterized by one parameter $\alpha$, defined by
$$\textrm{frequency of $f$} \propto \Beta(f+1-\alpha, \alpha).\label{eq:evert}$$

It is involved in our analysis for the following reason:
\begin{proposition}\label{proposition:evert}
Suppose that we draw samples from a discrete Zipf-Mandelbrot distribution with parameters $\gamma$ and $t$.
If the number of samples is large, then for all small $f$, the expected number of $f$-legomena is proportional to $\Beta(f+1-\alpha, \alpha)$ where $\alpha = 1+1/\gamma$.
\end{proposition}

Paraphrasing, this says that the distribution-on-frequencies for a discrete Zipf-Mandelbrot distribution is the Evert distribution. This result was obtained by Evert~\cite{Evert04} though he expressed it in terms of ``type density functions.'' We re-prove it in \prettyref{app:b}.

We remark that the proof of \prettyref{proposition:evert} remains valid even if the discrete Zipf-Mandelbrot distribution is perturbed by altering some of the highest probabilities, which ensures that it is still valid even if outliers \`a la \prettyref{sec:zmd} occur.

\subsubsection{Remarks}
In~\cite{Newman05,CSN09}, the focus of the analysis is on continuous power law distributions, and for that, the analogue of \prettyref{proposition:evert} is to use a simple power law with exponent $\alpha$ instead of an Evert distribution. Though the Evert distribution is not mentioned in~\cite{Newman05}, it is remarked that the another distribution, the Yule distribution, is an ``an alternative and often more convenient form'' of the discrete power law. These conveniences are mathematical in nature: the normalizing constant, expectation, variance, and moments of the Yule distribution have nicer closed forms than a pure power law. (And it is reasonable to use in power law analysis because up to a scaling factor, the Yule distribution becomes a discrete power law in the limit.)
These conveniences holds for the Evert distribution too, since Evert and Yule differ only by a shift. For instance, our fitting code utilizes the identity
$\sum_{f=1}^{F_{\max}} \Beta(f+1-\alpha,\alpha) = \Beta(2-\alpha,\alpha-1) - \Beta(F_{\max}+2-\alpha,\alpha-1).$

\subsection{Maximum Likelihood}
The Evert distribution allows us to compute the most likely value of $\alpha$ for the collection of frequencies $\vF.$ Writing $E^\alpha_f$ for $\Beta(f+1-\alpha, \alpha)$, and $C^\alpha$ for the normalizing constant with $C^\alpha \cdot \sum_{f=1}^{F_{\max}} E^\alpha_f = 1$, we seek the $\alpha$ that maximizes $$\prod_{k=1}^M C^\alpha \cdot E^\alpha_{F_k}.$$
This can be determined numerically using binary search, using logarithms since the numbers involved are very small. Then we determine the parameter $\gamma$ using $\gamma = 1/(\alpha-1).$

The only remaining issue is how to determine the value of the shift parameter $t$ that has maximum likelihood. \prettyref{proposition:evert} does not help since $t$ plays no role in its conclusion. (The reason for this apparent paradox is that the approximation guarantee of \prettyref{proposition:evert} is only valid for small frequencies.) Nonetheless, we can determine a value for $t$ using some ideas from the analysis of the continuous case~\cite{Newman05,CSN09}.\footnote{The authors of \cite{Newman05,CSN09} note that the continuous model reasonably resembles the discrete model when thinking about larger frequencies; the smallest continuous variables are the ones that would have to be distorted the most in order to become quantized. Thus, the inaccuracies of \prettyref{proposition:t} are complementary to those of \prettyref{proposition:evert}.}

\begin{proposition}\label{proposition:t}
Let $\alpha = 1+1/\gamma$. Suppose we draw a sample from the bounded continuous power-law distribution with exponent $-\alpha$ and domain $(1, (\frac{t}{t+M+1})^{-\gamma})$.
Then the $(M+1)$-quantiles of this random variable are proportional to $(t+1)^{-\gamma}, (t+2)^{-\gamma}, \dotsc, (t+M)^{-\gamma}$. Furthermore, the choice of $t$ that maximizes the likelihood of observing $\vF$ is $t = (M+1)/(F_{\max}^{1/\gamma}-1)$.
\end{proposition}

The first conclusion says that a ``typical'' draw of $M$ items from this continuous distribution-on-``frequencies'' is a model for the Zipf-Mandelbrot distribution. The second conclusion gives us the rule that we use to compute $t$ in our statistical fitting. We prove the proposition in \prettyref{app:t}.

\section{Fitting the Data}
In Evert's paper \cite{Evert04}, rather than using maximum-likelihood, he proposes estimating $\alpha$ using a Chi-squared test on the first few $\#F^{-1}(1), \#F^{-1}(2), \dotsc$ values. This approach is implemented by the R library \verb|zipfR| of Evert and Baroni~\cite{EB07}.


\begin{table}[htb!]
$$
\begin{array}{|c|c|c|}
\hline
& \mathrm{Max.~likelihood} & \chi^2 ~ \mathrm{ min.} \\
\hline
\mathrm{Java} & \alpha=1.216, t=33.1 & \alpha=1.225, t=25.7 \\
\hline
\mathrm{Python} & \alpha=1.165, t=44.7 & \alpha=1.143, t=65.7 \\
\hline
\mathrm{Python}^\mathrm{w/o} & \alpha=1.131, t=99.8 & \alpha=1.133, t=92.9 \\
\hline
\end{array}
$$
\caption{Results of fitting our data sets to Zipf-Mandelbrot distributions with both methods. Python$^\mathrm{w/o}$ indicates the Python data set
with the 3 commonest messages removed.}
\label{table:fits}
\end{table}

\begin{figure}[htb!]
\centering
\includegraphics[width=6cm]{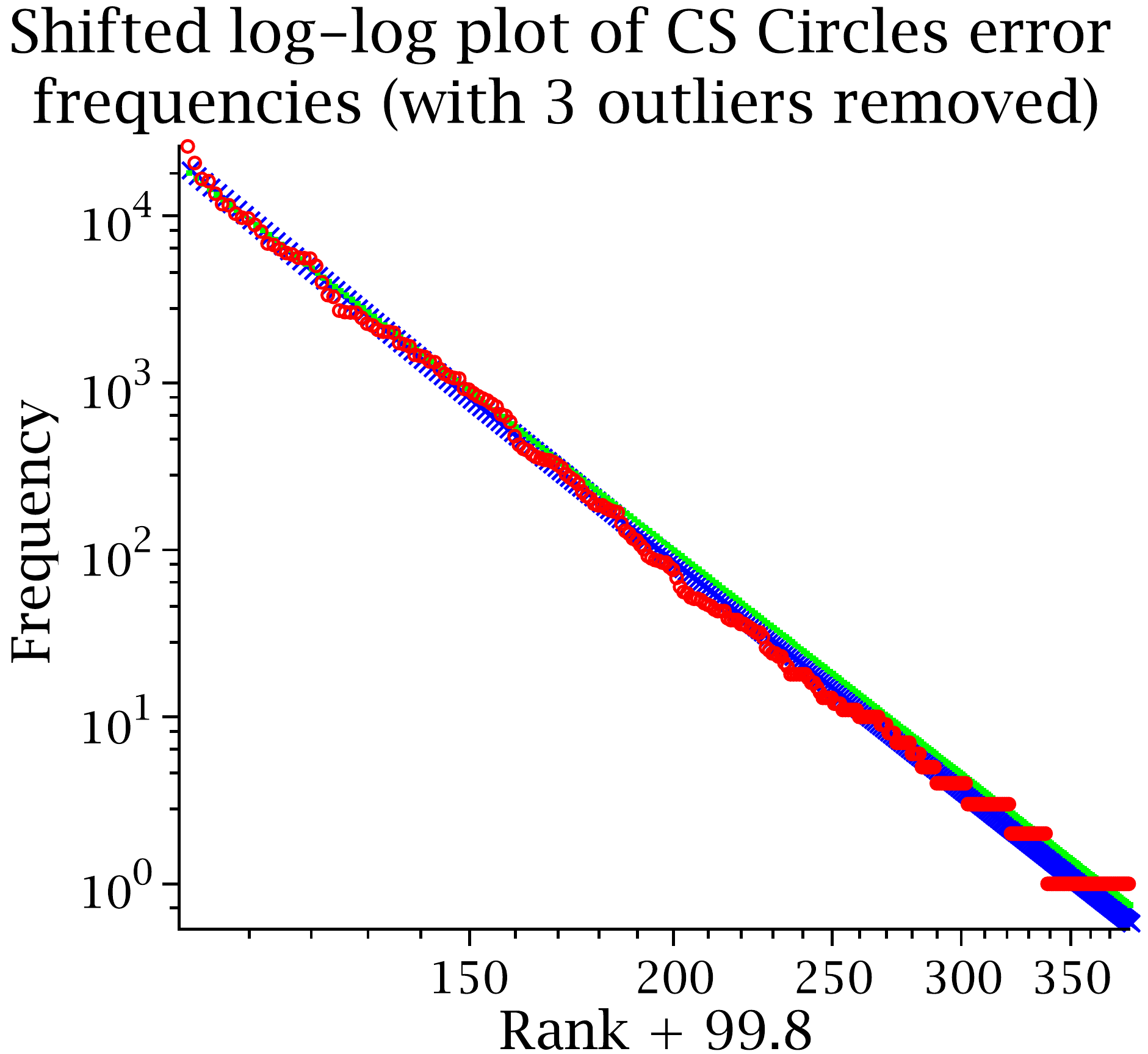}
\caption{Plot of the Python data set on shifted log-log axes, with shift $t$ from maximum likelihood estimation. Red: observed frequencies; blue: Chi-squared fitted Zipf-Mandelbrot distribution; green: maximum likelihood fitted Zipf-Mandelbrot distribution.}
\label{fig:finalcsc}
\end{figure}

\begin{figure}[htb!]
\centering
\includegraphics[width=6cm]{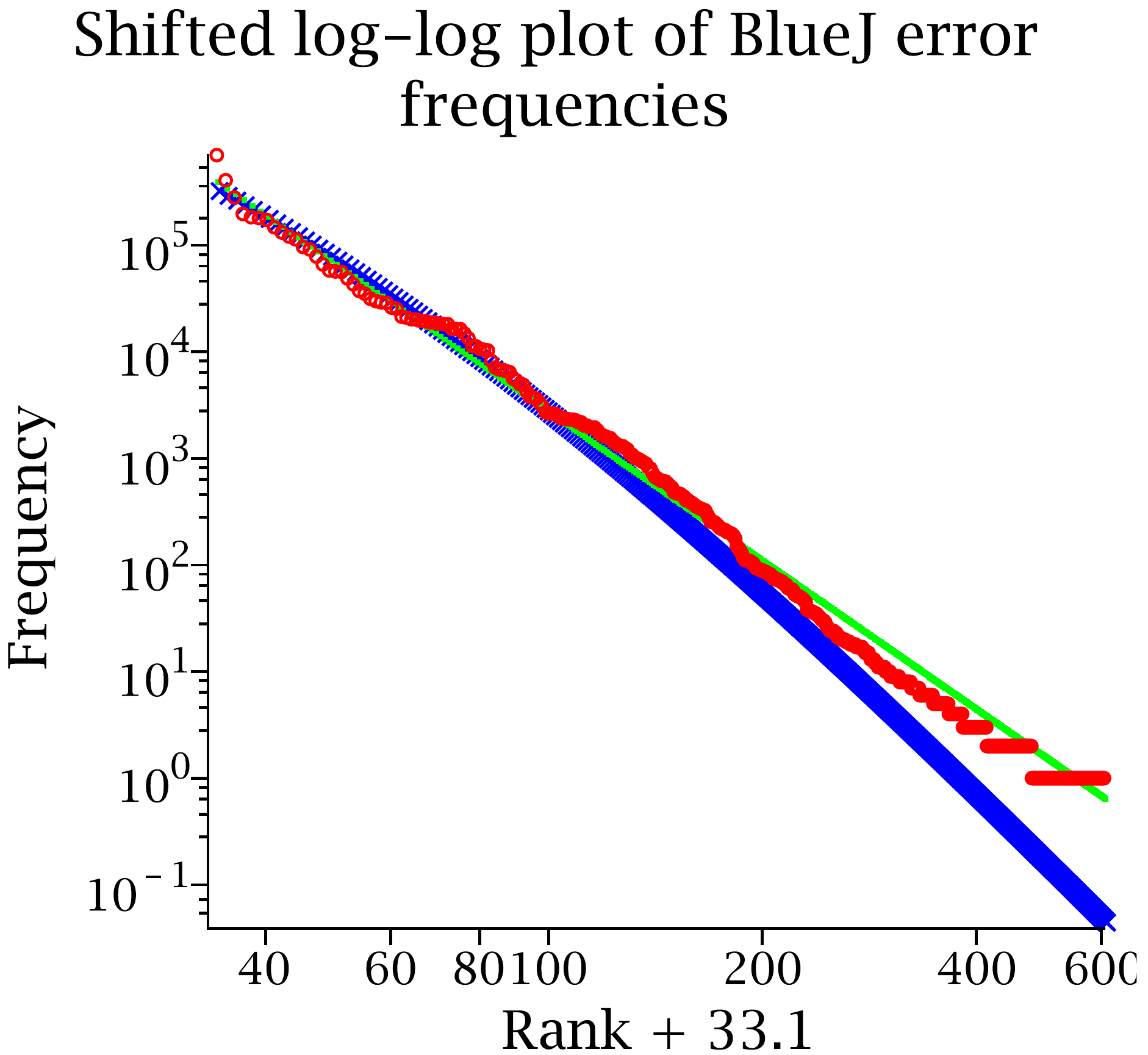}
\caption{Plot of the Java data set, analogous to Figure~\ref{fig:finalcsc}.}
\label{fig:finalj}
\end{figure}

We fit our data sets to the Zipf-Mandelbrot family of distributions, using both the Chi-squared approach, and the maximum-likelihood method of Propositions \ref{proposition:evert} and \ref{proposition:t} (implemented in Maple).
The results of the fitting are shown in Table~\ref{table:fits}. The fit for the Python data set improved greatly by treating the three most common errors as outliers (c.f.~Figure~\ref{fig:rawdata}). In Figures~\ref{fig:finalcsc} and~\ref{fig:finalj} we show shifted log-log plots of the observed fits (the Python plot omits the outliers). 
For the Python-without-outliers data set, both methods give a good fit. For the Java data set, the maximum-likelihood method gives a significantly better fit than the Chi-squared method.

\section{Future Work}
A few very short questions for future work are:
(1) can the good fit be replicated in other Java/Python systems?
(2) if so, what properties of the user base or programming ecosystem affect the $\alpha$ and $t$ parameters?
(3) do error messages in other languages also follow a Zipf-Mandelbrot distibution?

In the context of the hypothetical extreme languages of Section~\ref{sec:related}, Python's slightly smaller value of $\alpha$ suggests that it tends to give more distinctive error messages. Is it actually giving more information in its errors? Could it alternatively be explained due to artefacts like the non-parallelism mentioned in Section~\ref{sec:datasets}?

It would be interesting to re-analyze the discrete data sets in \cite{Newman05} using the Evert maximum likelihood method. Specifically, this could be done for the data sets for word frequency, web hits, telephone calls, and citations, which are discrete distributions coming from a population that is large enough to be effectively infinite. Additionally, it would be interesting to apply the Kolmogorov-Smirnov test suggested in \cite{Newman05} to the Evert maximum likelihood method, to be more rigorous in our approach.



From a more practical perspective, it would be not hard, and of a great potential benefit, to release a systematic data set of good beginner-friendly explanations of the top errors in different programming languages. Further work could try to quantify if this improves the ability of beginner students to program independently.

\subsection*{Acknowledgment}
We thank the SIGCSE Special Projects committee, whose Summer 2013 grant for CS Circles provided funding when this work was initiated~\cite{PGV15}, and the Blackbox project for their work on providing accessible huge data sets. We thank undergraduate assistants Ayomikun (George) Okeowo and Pallavi Koppol for their work on sanitizing and writing explanations. Thanks to Jurgis Pralgauskis for translating the Python explanations to Lithuanian. We also thank the PLATEAU referees for their helpful suggestions.

\bibliographystyle{abbrvnat}
\bibliography{biblio}

\appendix

\section{Proof of \prettyref{proposition:evert}}\label{app:b}
Fix a constant $f$ and consider $N$ as growing to infinity. By linearity of expectation, the expected number $\E[\#F^{-1}(f)]$ of $f$-legomena is equal to the sum, over all $\ell$, of the probability that word $\ell$ occurs exactly $f$ times in our sample. For large $N$, any word with frequency bigger than a constant has vanishingly small probability of occurring only $f$ times. So for a word that may become an $f$-legomenon, its number of occurrences is well-approximated by a Poisson random variable, since it is a sum of many Bernoulli random variables, each with a small individual expectation. The expected number of occurrences of the $\ell$-th most common word is $NC(\ell+t)^{-\gamma}$, so the number of times we observe it is a Poisson variable with expectation $NC(\ell+t)^{-\gamma}$.

This means that for any constant $f$, by the definition of a Poisson variable,
$$\Pr[\textrm{word $\ell$ appears exactly $f$ times}] = \frac{(NC(\ell+t)^{-\gamma})^f}{f!\exp(NC(\ell+t)^{-\gamma})}.$$
Thus, the expected number of words appearing $f$ times is
$$\E[\#F^{-1}(f)] = \sum_{\ell=1}^\infty \frac{(NC(\ell+t)^{-\gamma})^f}{f!\exp(NC(\ell+t)^{-\gamma})}.$$
We approximate this infinite sum with the infinite integral
$$\E[\#F^{-1}(f)]=\int_{1}^\infty \frac{(NC(x+t)^{-\gamma})^f}{f!\exp(NC(x+t)^{-\gamma})}dx.$$
To evaluate it, we substitute $y = NC(x+t)^{-\gamma}$, i.e.~$x = (\frac{y}{NC})^{-1/\gamma}-t$ and so $dx = -\frac{1}{\gamma}(CN)^{1/\gamma}y^{-1-1/\gamma} dy$, giving
$$\E[\#F^{-1}(f)] = \frac{(CN)^{1/\gamma}}{f! \gamma} \int_0^{NC^{-1}t^{-\gamma}} \frac{y^{f-\frac{1}{\gamma}-1}}{\mathrm{e}^y} dy.$$
Again assuming $N$ large, the above integral is well-approximated by replacing the upper bound by $+\infty$. Therefore, taking the terms that do not depend on $f$ into the  constant of proportionality, we find that
\begin{align*}
\E[\#F^{-1}(f)] &\propto \frac{1}{f!} \int_0^{+\infty} \frac{y^{f-\frac{1}{\gamma}-1}}{\mathrm{e}^y} dy
\\ &= \frac{\Gamma(f-1/\gamma)}{\Gamma(f+1)} \\ &\propto \Beta(f-1/\gamma, 1+1/\gamma) = \Beta(f+1-\alpha, \alpha).
\end{align*}

\section{Proof of \prettyref{proposition:t}}\label{app:t}

Let $U(a, b)$ denote a random variable from the uniform distribution on $(a, b).$ Our starting observation is that the continuous power-law distribution with exponent $-\alpha$ and unbounded domain $(1, +\infty)$ is identical in distribution to $U(0, 1)^{-\gamma}$. See, for instance, \cite[App.~D]{CSN09}.

Therefore, adding the bound to get the continuous power-law in the hypothesis of the theorem, said distribution is identical in distribution to $U(\frac{t}{t+M+1}, 1)^{-\gamma}$.

The $(M+1)$-quantiles of $U(\frac{t}{t+M+1}, 1)^{-\gamma}$ are $((t+k)/(t+M+1))^{-\gamma}$ for $k=1, \dotsc, M$, so the first conclusion follows.

Finally, the smaller the domain $(1, (\frac{t}{t+M+1})^{-\gamma})$, the larger the probability density function at the observed $\vF$ values, except that we need $(\frac{t}{t+M+1})^{-\gamma} \geq F_{\max}$ for $F_{\max}$ to be observable at all. This proves the second conclusion.

%
%
%

\end{document}